\documentclass[aps,preprintnumbers,nofootinbib,superscriptaddress,twocolumn]{revtex4}

\usepackage{graphicx}
\usepackage{epsfig,graphics}
\usepackage{amsfonts,amssymb,amsmath}
\usepackage{amsthm}

\newcommand{\comment}[1]{}

\theoremstyle{plain}

\theoremstyle{definition}

\begin{document}

\title{A No-summoning theorem in Relativistic Quantum Theory}

\author{Adrian \surname{Kent}}
\affiliation{Centre for Quantum Information and Foundations, DAMTP, Centre for
  Mathematical Sciences, University of Cambridge, Wilberforce Road,
  Cambridge, CB3 0WA, U.K.}
\affiliation{Perimeter Institute for Theoretical Physics, 31 Caroline Street North, Waterloo, ON N2L 2Y5, Canada.}

\date{January 25, 2011; revised April 4, 2011 (minor)}

\begin{abstract}
Alice gives Bob an unknown localized physical state at some point P.
At some point Q in the causal future of P, Alice will ask Bob 
for the state back.  Bob knows this, but does
not know at which point Q until the request is made. 
Bob can satisfy Alice's summons, with arbitrarily short delay, for a quantum state
in Galilean space-time or a classical state in Minkowski space-time. 
However, given an unknown quantum state in Minkowski space-time, he
cannot generally fulfil her summons.  This {\it no-summoning theorem} is a
fundamental feature of, and intrinsic to, relativistic quantum theory.
It follows from the no-signalling principle and the no-cloning theorem, but not from either alone.   
\end{abstract}

\maketitle

\section{Introduction}
\label{sec:intro}
Although some of the fundamental properties of quantum theory -- for
example, the superposition principle -- were understood very early,
other key insights were made only later.   Quantum entanglement was first
described by Schr\"odinger \cite{entanglement} only in the 1930s; it was not till the 1960s that
Bell showed that quantum theory violates local causality \cite{belllocbeables,bellloccaus,bellbook,CHSH};  
some other important aspects of the delicate relationship between the general quantum measurement postulates and the
no-signalling principle were not completely understood until
even more recently\cite{gisinone,gisintwo,czachor,aknlwithouts,infocaus}.   
These and other features of quantum theory have been greatly
illuminated in recent decades, with the development of quantum
computing, quantum cryptography and quantum information theory, 
which have inspired a perspective on quantum theory in terms
of tasks and resources involving physical information.   

Conversely, considering
which quantum tasks are possible or impossible has led to 
significant discoveries in quantum communication (e.g. \cite{teleportation})
and quantum
cryptography (e.g. \cite{wz,dieks,mayersprl,lochauprl}).   
One of the earliest, simplest and most celebrated results in this field was
the quantum no-cloning theorem\cite{wz,dieks}, whose proof is mathematically trivial,
but which nonetheless encapsulates a fundamental fact about quantum theory 
that had previously lain unremarked.   The no-cloning theorem inspired
several other significant results, including independent proofs of the
impossibility of determining an unknown quantum state\cite{day} and the impossibility of distinguishing
between non-orthogonal states\cite{yuen}, the no-deletion theorem\cite{braunsteinpati}, 
the no-broadcasting theorem for mixed states\cite{nobroadcast}, a general no-cloning
theorem incorporating several of these results\cite{lindblad}
and a proof that it is impossible to clone with partial ancillary information\cite{jozsa}.   
A further significant extension was the introduction of the idea of
partial fidelity cloning\cite{bh}, and the discovery of universal algorithms for attaining the 
best possible state-independent fidelities for $M \rightarrow N$ partial cloning \cite{gisinmassar,bem,werner,keylwerner}.   
The recent discovery of the principle of information causality\cite{infocaus} has given
further interesting insight into quantum theory and its relationship with special relativity. 

All of the results mentioned thus far shed interesting light on the relationship
between quantum theory and special relativity, and several of them are crucial
to our current understanding.   However, they all describe features already
evident in non-relativistic quantum mechanics.
Since, as we currently understand things, relativistic quantum 
theory is closer to the true description of nature than quantum mechanics, 
there remains a compelling motivation to understand which
properties and principles are intrinsic to relativistic quantum
theory.   The motivation is even more compelling since 
we understand relativistic quantum theory so poorly compared to
quantum mechanics.  We have an informal intuitive understanding of
many features of Lorentz invariant quantum field theories with local
interactions, but as yet no rigorous definition of any non-trivial
relativistic quantum theory.  One might hope ultimately to supply
such a definition by identifying the principles that such a theory must
satisfy.

This paper describes a simple new task -- {\it summoning} an
unknown localized physical state -- which, unlike 
those mentioned above, distinguishes non-relativistic quantum mechanics
and relativistic classical physics from relativistic quantum theory.    
As we show below, an unknown state can be successfully summoned in 
the first two theories, but not in the last.   

Like the no-cloning
theorem, to which it is closely related, and several of the other 
results mentioned above, the no-summoning theorem is relatively easy
to prove once stated.  It nonetheless
encapsulates a significant and previously unremarked feature of 
relativistic quantum theory and (again like the no-cloning theorem,
though more directly)
leads to some significant and qualitatively new cryptographic
applications.   

It is also worth pointing out that, while ultimately mathematically
simple, the result is not quite
as transparent as naive intuition might suggest.   
Quantum theory allows the information in an initially localized
quantum state to be delocalized, or ambiguously located, in
quite counter-intuitive ways.  In particular, the possibility
of quantum teleportation means that, in a certain sense, an 
unknown initial state can be reconstructed in different ways,
from different subsystems, at different and spacelike separated
locations.  One cannot thus argue that a state must follow
some definite timelike or null path in spacetime: indeed, it is
precisely the failure of this reasoning that led recently to the
breaking (\cite{kms}; see also \cite{buhrmanetal}, where this attack
is generalized using earlier work of Vaidman \cite{vaidman}) of purportedly
secure protocols \cite{chandranetal,malaney,malaneyarxiv} for quantum tagging
\cite{taggingpatent,kms} (also
called quantum position authentication \cite{chandranetal}).  Another motivation for 
the present result is to give a precise and positive statement about
the localizability of an unknown state, to set alongside the
intriguing negative results illustrated by Refs. \cite{vaidman,kms,buhrmanetal}.  

\section{Summoning a state: idealized scenario}

We initially suppose that space-time is either Galilean or
Minkowski and that nature is described either by classical mechanics
or quantum theory: we will distinguish these cases below.   
We suppose that both parties have arbitrarily efficient technology, limited
only by the relevant causal structure and physical theory.  In particular, to simplify
the discussion initially, we suppose that Alice and Bob can independently and securely access 
any relevant point in space-time and instantaneously process and
exchange information there.  We also suppose that their preparations,
information processing, 
communications and measurements
are error-free and have unbounded precision and capacity.   
We will relax these idealized and somewhat unphysical assumptions later.   

Consider two agencies, Alice and Bob, who agree in advance on some
space-time point $P$ that they can access independently. 
Idealizing, we suppose they are independently able to access 
every point in the causal future of $P$, and that each is able to keep
information everywhere secure from the other party unless and until they
choose to disclose it.\footnote{
In a more realistic model, which deviates from our idealized scenario
but can still
illustrate all the key features of our discussion,
Alice and Bob could both be large independent collaborative groups of people,
with each group having their own independent secure network of quantum
devices distributed throughout a region.} 
At the point $P$, Alice gives Bob a localized physical system $S$ (which in
the ideal case we treat as pointlike, and which we will assume 
has zero internal Hamiltonian and can, if Bob wishes, easily be kept isolated with
no interaction with the environment) whose state is known to her
but not known in advance to Bob.   For definiteness, let us say that
the state is drawn randomly from some probability distribution $\mu$ 
known to them both.   Until he receives $S$, Bob has no more information about $S$ than is
implied by $\mu$.  Once he receives $S$, he has the additional information carried by $S$
(but nothing further).   

It is agreed that at some point $Q$ in the 
causal future of $P$, to be decided by Alice but not known in advance
by Bob, Alice will ask Bob to return $S$ in its original 
state.   Bob is required to do this in a way that identifies the
returned system $S$, by specifying the relevant classical or quantum
degrees of freedom.\footnote{He may not give Alice some larger system --
for instance, a collection of all possible states -- 
and then argue that he has satisfied the summons because
he has indeed supplied $S$, albeit in the form of an unspecified subsystem.}

Again, for definiteness, let us say that Alice chooses $Q$
randomly from some probability distribution $\mu'$ known to both parties. 
At any point $Q'$ that does not lie in the causal future of the chosen $Q$,
Bob has no more information about the choice of $Q$ than is implied by 
$\mu'$ and the fact that $Q$ is not in the causal past of $Q'$. 
We say that Alice {\it summons} the state from Bob at point $Q$, and
that Bob {\it complies} with the summons if he returns an identified
verifiable copy of the system $S$ to Alice at $Q$.   
Here a {\it verifiable copy} is defined operationally, as a state that Alice has zero
probability of distinguishing from $S$ by any physical means.\footnote{We discuss
the case of probabilistic discrimination, which is relevant when we
consider quantum theory, below.}

\section{Summoning in classical theories and in non-relativistic quantum mechanics}

It is easy to see that, in our idealized scenario, Bob can satisfy Alice's request in 
classical mechanics, in either Galilean or Minkowski space-time.    Once he receives
the state at $P$, he immediately measures it to infinite precision, and broadcasts
the result to all points in the causal future of $P$.    When he receives a summons
at $Q$, he instantaneously receives and processes this broadcast, making a 
perfect copy of $S$, which he returns to Alice. 

The same is true of quantum mechanics in Galilean space-time.   For example, if the
point $P$ has coordinates $(x_P , t_P )$ in some inertial frame, Bob can simply hold the state $S$
at position $x_P$ until he receives a summons at $Q = (x_Q , t_Q )$.    He then sends an
instantaneous signal from $Q$ to $(x_P, t_Q)$, on receipt of which he instantaneously 
sends the state $S$ to $Q$ and returns it to Alice.   

\section{No summoning in relativistic quantum theory}

\begin{quote}
GLENDOWER  I can call spirits from the vasty deep.

HOTSPUR Why, so can I, or so can any man;
But will they come when you do call for them?\footnote{As Hotspur notes, strictly speaking
it would be more precise to speak of no successful summoning, or of the impossibility of general
compliance with summonses.  However we side with Glendower here in preferring succinctness to perfect
precision.}  

(Wm. Shakespeare, {\em Henry IV, Part 1})  
\end{quote}  

To fix a definite example, take the probability distribution $\mu$ to be the uniform distribution
over pure quantum states in some dimension $d \geq 2$, and suppose that at point $P = ( x_P , t_P )$
Alice gives Bob an unknown pure state $\rho$ drawn from this distribution.  
Suppose that the point $Q$ is drawn from a probability distribution
$\mu'$ uniform over the set $X$ of 
points that have time coordinate $t_Q = t_P + t$ in some given inertial
frame, for some fixed $t > 0$, and that are lightlike 
separated from $P$; i.e. $X$ is a spatial sphere with centre $x_P$
and radius $t$, at time $t_Q$.   
Each of these points is spacelike separated from all of the others, so
Bob learns no extra information about the choice of $Q$ at any point in
the causal past of any possible $Q$.  

Bob may instantaneously carry out any quantum operations on $\rho$, together with an ancilla, 
at $P$, and then send outputs at light speed to his representatives at any or all of the possible 
points $Q$.\footnote{He could also carry out further operations at points $P$
on light rays between $P$ and any possible $Q$, but since he learns nothing new at these points, this
gives him no advantage.   He could also send communications along timelike paths
from $P$, along timelike paths from any $P'$ between $P$ and any
possible $Q$, or along lightlike paths from such a $P'$ in a direction other than
towards the relevant possible $Q$: again, he gains no advantage, since these communications
will not reach any of the possible $Q$.}
Bob chooses and follows some strategy, with the result that at each
possible $Q$ he has some (not necessarily pure) state $\rho_Q$ which
he will hand over to Alice in response to the summons, if it arrives
at $Q$.\footnote{Again, he could plan to carry out further instantaneous information
  processing at $Q$ if the summons arrives there, but this gives him
  no advantage: he loses nothing by pre-preparing all the states at
  $P$.}

Fix now on two possible summoning points, $Q_0$ and $Q_1$.   If the
summons arrives at $Q_i$, then Bob successfully complies only if 
$\rho = \rho_{Q_i}$.   However, the no-signalling principle
means that the state $\rho_{Q_1}$ must be independent of whether
or not Alice chooses to summons at $Q_0$, and vice versa. 
So Bob can successfully comply with both possible summonses only
if $\rho_{Q_0} = \rho_{Q_1} = \rho$.   But this would imply
that, on the spacelike surface defined by time coordinate $t_Q$,
Bob has made two perfect copies of $\rho$, which contradicts
the no-cloning theorem.   Summoning an unknown state is thus not
generally possible in relativistic quantum theory.

\subsection{The non-ideal case: finite processing and response time} 

At this point, it is natural to ask whether the no-summoning theorem could be
an artefact of some of the idealizations in our model, and so
to consider more realistic models.   We start by considering ways of relaxing the 
assumptions that Bob should be able to carry out information processing
instantaneously, communicate at light speed, or respond instaneously
to a summons.   

One way of doing this is to allow finite time margins of error, in a given
fixed inertial 
frame, for each of these processes.  
In the concrete example considered above, we can allow Bob some time
for information processing after receiving $\rho$ at $P$, and 
also some margin for slower than light communication, by taking 
the set of possible $Q$ to be the set of points $X'$ with time
coordinate $t_Q= t_P + t$ (as before) and some fixed timelike separation
$t'$ from P, where $0 < t' < t$, i.e., the set of points in $X'$ form a spatial sphere with centre $x_P$
and radius $r = ( t^2 - (t' )^2 )^{1/2}$, at time $t_Q$.  
We can then allow Bob a further finite time $\delta$ to process
information and respond to the summons after it is received at $Q =
(x_Q, t_Q )$, so that Bob is now required to return the state at the spacetime 
point $Q^\delta = (x_Q, t_Q + \delta)$.   We require $ \delta \ll r$, so that for
most pairs $(Q_0 , Q_1)$ of points with $Q_i \in X'$ the causal pasts of
$Q^\delta_0$ and $Q^\delta_1$ do not intersect at time $t_Q$.   

Now let $(Q_0 , Q_1)$ be a pair of summoning points with this last
property --- for example,
a pair of antipodes on the sphere defining $X'$.   
If Bob receives a summons at $Q_0$, he can successfully comply only if
he can produce the state $\rho$ by quantum operations on
the distributed state $\phi_0$ available to him in the spatial region
lying in the causal past 
of $Q^\delta_0$ at time $t_Q$.\footnote{This is a necessary but not sufficient condition:
in fact, he has no strategy which allows him to propagate even these distributed states to 
all possible returning points $Q^\delta$.}  Similarly, if he receives 
 a summons at $Q_1$, he can successfully comply only if
he can produce the state $\rho$ by quantum operations on
the distributed state $\phi_1$ available to him in the spatial region
lying in the causal past 
of $Q^\delta_1$ at time $t_Q$.   But the states $\phi_0$ and $\phi_1$
belong to disjoint factors of the space of states in Bob's control
at time $t_Q$, since they are localized in disjoint regions.  
Hence our previous argument applies.  The state $\phi_0$ is
independent of whether Alice chooses to summon at $Q_1$, and vice
versa, by the no-signalling principle.   Hence Bob is in a 
position to comply with a summons at $Q_0$ or (alternatively) a
summons at $Q_1$ only if he can separately create copies of $\rho$ 
from both $\phi_0$ and $\phi_1$, which again violates the no-cloning
theorem.

We have expressed Bob's allowed response time in a fixed frame here,
as it makes his constraints easy to visualize and understand.
The frame choice is defined by the set of allowed summoning points.
This breaking of Lorentz invariance is necessary in any non-trivial practical example -- 
in practice the set of summoning points will always be compact and so not
Lorentz invariant, and the parties' devices will define preferred sets
of coordinates.   Abstractly, though, it is interesting to consider 
Lorentz invariant versions of summoning incorporating time delays.
One way of doing this is to define the set of allowed summoning points
to be all points $Q$ in the future light cone of $P$ such that
$\tau(P,Q) = t'$, the set of allowed response points to be points 
$R$ in the future light cone such that $\tau(P,R) = t' + \delta$, and
require that $R$ is determined, once $Q$ is fixed, by the constraint
that $P$, $Q$ and $R$ are colinear.  Here we require
$\delta \ll t'$.   

\subsection{Relation to standard mistrustful cryptography}

Note that relaxing our idealized assumptions allows us to fit the task 
of summoning into the standard cryptographic model \cite{kentrelfinite}  for mistrustful 
parties in Minkowski space-time.   The non-idealized case allows us to
drop the assumption that Alice and Bob each have independent secure
access to every space-time point.   Instead, we can suppose that Alice and Bob
control suitably configured disjoint regions of space-time, their ``laboratories''. 
Each trusts the security of their laboratory and all devices contained
within it, but need not trust anything outside their laboratory.   

We can take Bob's laboratory to be a  
connected region of space-time that includes, near its boundary, $P$ and all allowed
summoning points $Q_i$, and includes line segments joining $P$ to each
$Q_i$: this allows Bob to receive a state at $P$ and transmit it
securely to any $Q_i$.    We can take Alice's laboratory to be a disjoint connected region
of space-time that includes a point $P'$ in the near causal past of
$P$, from which she sends the unknown state to $P$, points $Q'_i$ in the
near casual past of each possible summoning point $Q_i$, from which she
sends a summoning request to the relevant $Q_i$, and points $Q''_i$ in
the near causal future of each summoning point $Q_i$, to which Bob is 
supposed to send the summoned state if the summons arrives at $Q_i$. 
This allows Alice to generate the unknown state securely, transmit it
to $P$, generate a summoning request securely (so that Bob cannot
predict it in advance), transmit it to $Q_i$, receive the summoned
state at $Q''_i$, and test it securely. 

As is standard in mistrustful cryptographic scenarios, we
assume that Alice and Bob are the only relevant parties -- no one else
is trying to interfere with their communications -- and that they 
have classical and quantum channels (which in principle can be made
arbitrarily close to error-free) allowing them
to send classical and quantum signals between the relevant points.   

\subsection{No approximately successful summoning} 

Another idealization in the model above is that we take the state $\rho$ to be localized at a point. 
We could, instead, consider the state to be localized in a finite region -- i.e. for its wave function to 
be zero outside the region.   This case can easily be handled as above, with a few more epsilonics.
However, this is still an idealization: physical states are only approximately localized.
So, if Bob and Alice treat the system as perfectly localized in a finite region, they will introduce
some probability of error in the verification step.   
A realistic model would also allow for errors in Alice's preparation and measurement.
Independently of these points, even in an ideal model, it is also interesting to ask whether 
Bob can generally respond to summonses in such a way that he will pass Alice's verification
test with a probability $p$ that can be made arbitrarily close to $1$.  

We are thus motivated to consider the possibility of approximately successful summoning.
We say that Bob can guarantee {\it $p$-compliance} with the summons if he has a 
strategy which, for each possible summoning point $Q$, allows him to generate a 
state ${\rho_Q}$ that he can return to Alice at (or, in the non-ideal case, appropriately near) $Q$, 
such that $ {\rm Tr} ( \rho_Q \rho ) \geq p$ for all $Q$.  
We can show that this is also not generally possible, for $p$ arbitrarily close to $1$, by generalizing the 
arguments above, using the bounds
\cite{gisinmassar,bem,werner,keylwerner,bbh,cerf,ffc,iblisdiretal,iag} on the fidelity of 
$1 \rightarrow 2$ quantum cloning for qudits.\footnote{See
  Ref. \cite{bcsummoning} for a quantitative discussion.}

\section{Discussion}

Summoning is most simply and naturally defined in the idealized case, but also 
has simple natural extensions to realistic models of information processing and 
to the probabilistic case.   It may be the first significant example of a simple information
theoretic task that essentially distinguishes relativistic quantum theory from non-relativistic
quantum theory and from relativistic classical physics.   Its impossibility has significant
implications for quantum computing and quantum cryptography, which deserve
further exploration.   

We have framed the task in terms of two parties, Alice, who creates and
knows the physical state, and Bob, who knows nothing about the state until he receives
it, and then only the information conveyed by the state itself.  
One could, alternatively, suppose that the state is not known to either Alice or Bob,
but obtained by Alice from a third party or from nature.   In this case, Alice cannot verify 
whether or not Bob (approximately) complies with the summons, and so we cannot define
the task operationally.   However, we can still define
success depending on whether the fidelity of the returned state to the original is
(close to) one.    In another version of this picture, one could instead think of nature as one of the
parties: for example, Bob could find a state in nature, and be required to produce
it at a summoning point that depends on some natural event which he cannot 
predict in advance.   

We have given examples to show that summoning is not generally possible in relativistic
quantum theory.   In essence, this follows from the fact that summoning is not possible
even in the simple case in which Bob knows that Alice will summon at one of two spacelike
separated points, or, in the non-ideal case, within one of two spacelike separated regions.  
Of course, one can construct examples in which summoning {\it is} 
possible: for example, if all the possible summoning points lie on a timelike curve in
the causal future of $P$.    It is not hard (if perhaps not always very illuminating) to list 
conditions that ensure that summoning is impossible in a particular example, for
either the ideal or non-ideal cases.   

Summoning is closely related to cloning, and we can extend the parallel to define other related impossible
tasks in relativistic quantum theory, such as summoning an unknown member of a 
set of non-orthogonal states, summoning an unknown state given partial
ancillary information, or summoning $N>M$ returned copies of $M$ states (where the states may have been
supplied at different spacetime points, and the summonses may arrive at different spacetime points). 
One can also consider the probabilistic versions of these tasks.     The arguments above generalize straight away
to these cases, using the existing non-relativistic results, for the case where the states are supplied
at a single point and the possible summoning points form a suitable sphere, and more generally
when the summoning points or regions are spacelike separated. 
However, one could (if motivated) certainly construct more complicated examples where the interplay
between possible quantum dynamics and relativistic geometry is more complicated and requires 
non-trivial new analysis. 
One could also consider summoning in fixed background spacetimes other than Minkowski space:
clearly, again, if the causal structure has no pathologies, no-summoning theorems
hold here too.   
One might also ask whether some natural version of the no-summoning principle will hold
in whatever theory turns out to unify quantum theory and gravity.   A negative answer would be
surprising.   

The most immediate and striking application of the no-summoning
principle, however, is to quantum cryptography.   Bob can comply
with a summons {\it somewhere} in the future of $P$, if he knows
the summoning point in advance (i.e. already at $P$).  In particular,
he can comply if he is free to choose the summoning point in advance. 
However, in making this choice, he excludes the possibility of being
able to comply at other spacelike separated points.  In other words,
in a sense that can be made precise, he has to commit himself in
advance to some particular choice of point at which to return the
state, to ensure that he will be able to return it.  By devising
schemes in which this choice remains hidden from Alice until
the state is returned, we can use this essentially simple intuition
to solve the long-standing problem of finding an unconditionally
secure implementation of the cryptographic primitive of quantum bit commitment
\cite{bcsummoning}, and can also implement other interesting cryptographic 
tasks \cite{otsummoning} with unconditional security.   Precise
statements and descriptions of these results are given elsewhere \cite{bcsummoning,otsummoning}.  
 
\acknowledgments
This work was partially supported by an FQXi mini-grant and by Perimeter Institute for Theoretical
Physics. Research at Perimeter Institute is supported by the Government of Canada through Industry Canada and
by the Province of Ontario through the Ministry of Research and Innovation.


\end{document}